\documentclass[conference]{IEEEtran}

\usepackage{multicol}
\usepackage{url}
\usepackage{pifont}
\usepackage{todonotes}
\usepackage{amsfonts}
\usepackage{amsmath}
\usepackage{CJKutf8}
\usepackage{tikz-cd}

\newcommand{\M}{\mathcal{M}}
\newcommand{\X}{\mathcal{X}}
\newcommand{\PM}{\mathcal{P}}
\newcommand{\N}{\mathbb{N}}

\begin{document}
%
\title{How to advance general game playing artificial intelligence by player modelling}

\author{\IEEEauthorblockN{Benjamin Ultan Cowley}
\IEEEauthorblockA{BrainWork Research Centre\\
Finnish Institute of Occupational Health\\
POBox 40, Helsinki 00250, Finland\\
\\
Cognitive Brain Research Group\\
University of Helsinki, Finland\\
Email: ben.cowley@helsinki.fi}
}

\maketitle

\begin{abstract}
General game playing artificial intelligence has recently seen important advances due to the various techniques known as 'deep learning'. However the advances conceal equally important limitations in their reliance on: massive data sets; fortuitously constructed problems; and absence of any human-level complexity, including other human opponents. On the other hand, deep learning systems which do beat human champions, such as in Go, do not generalise well. The power of deep learning simultaneously exposes its weakness.
Given that deep learning is mostly clever reconfigurations of well-established methods, moving beyond the state of art calls for forward-thinking visionary solutions, not just more of the same.
I present the argument that general game playing artificial intelligence will require a generalised player model. This is because games are inherently human artefacts which therefore, as a class of problems, contain cases which require a human-style problem solving approach. I relate this argument to the performance of state of art general game playing agents. 
I then describe a concept for a formal category theoretic basis to a generalised player model. 
This formal model approach integrates my existing 'Behavlets' method for psychologically-derived player modelling:

Cowley, B., \& Charles, D. (2016). Behavlets: a Method for Practical Player Modelling using Psychology-Based Player Traits and Domain Specific Features. \textit{User Modeling and User-Adapted Interaction}, 26(2), 257–306.
\end{abstract}






\section{Introduction}
\label{intro}

\begin{CJK}{UTF8}{min}
	\textit{Atari (当たり, あたり, or アタリ) nominalized form of ataru (当たる, あたる, or アタル) (verb), meaning:}
\end{CJK}
\begin{center}
	\large{\textit{"to hit the target"}}
\end{center}

With great fanfare and much publicity, recent studies have claimed solutions to two important landmarks of game-playing artificial intelligence (AI) - competitive world-class Go \cite{Silver2016} and (constrained) generalised game playing \cite{Mnih2015} \footnote{Although see \cite{Schmidhuber2015} for criticism of the claimed originality.}. Each of these problems is an exemplar 'difficult problem' in AI, and each solution uses variations on the 'deep neural network' learning method. However despite this leap of progress, there remains the need to move beyond the constraints exploited by these deep learning methods.

Beyond the recent advances in game playing AI lie two still more daunting challenges: to build AI that can react to human players as individuals; and to play games of imperfect information and strategic scope \cite{Hassabis2016}. I argue that both of these problems are related, and therefore share a common solution. The link, or intersection in problem space, is that games are essentially \textit{human-relevant artefacts}. Encoding human-style playing strategies will enable not only better responses to human players, but better responses to problems of recreational interest to humans. Therefore both problems can be addressed with an improved understanding of human play, by building a \textit{generalised player model}. A general player model should be thought of primarily as a system for expressing the supra-specific elements that apply to all players, especially cognition, emotion and personality. The specific implementation then depends on the game.

As well as advancing the state of art, \cite{Silver2016,Mnih2015} demonstrate, directly and indirectly, that human-level performance (in specific problems) is different to human-level capability.
The senior author of both papers (they each come from Google's DeepMind lab) hinted at this \cite{Hassabis2016}:
\begin{quotation}
	I think for perfect information games, Go is the pinnacle... There are other games — no-limit poker is very difficult, multiplayer has its challenges because it’s an imperfect information game. And then there are obviously all sorts of video games that humans play way better than computers, like StarCraft... Strategy games require a high level of strategic capability in an imperfect information world — "partially observed," it’s called.
\end{quotation}

The specifics of the two systems nicely illustrate the problem. In both cases, their solution is largely a method to deal with very large possibility spaces, which were tackled by constraint of the search space and learning from large amounts of data, but in different ways.

\cite{Silver2016}'s 'alphaGo' system represented the Go board with a $19\times19$ matrix, and used Monte Carlo tree search (MCTS) enhanced by deep convolutional neural networks, to reduce the search space and determine play strategy. The algorithm first used supervised learning from existing games; then performed unsupervised reinforcement learning (RL) by playing against itself.
I term this the 'strategic approach', learning by deeply examining playing strategies (in the MCTS rollouts). It does not generalise well because large quantities of human-played games need to be used in the supervised part of training.

\cite{Mnih2015}'s 'Atari' system presented a deep RL algorithm with a $84\times84\times4$ map of pixels of the game screen from 49 Atari games. The algorithm then played a very large number of games to learn a policy. The performance was expressed proportional to a human expert, achieving $>$90\% in 26 games. At the other end of the scale, authors state that "\textit{temporally extended planning strategies still constitute a major challenge}".
Indeed, examining the lowest scoring games ($<$10\%) shows that they all share the features of an extensive game world, and the need to manage resources or tokens across this world. Thus the 'general approach' does not handle strategy well.
Especially in the two lowest scoring games, \textit{Montezuma's Revenge} and \textit{Private Eye}, the relationship between current task and overall possibility space is quite non-linear, as players must track back and forth through the game world to search, transport and use tokens such as keys. Compared to this, the strategy the authors actually describe their algorithm as having found in Breakout, a totally deterministic and perfect information game, is trivial. 

The poor performance (compared to a human) in this type of open problem is mirrored of the work of \cite{Acuna2010}. Given feedback, humans were shown to achieve close-to-optimal solutions on certain classes of computationally hard problems, e.g. Travelling Salesman, by heuristically exploiting structure in the 'typical instances'.

%
%
%
%
%

\cite{Marcus2016} has also pointed out how the DeepMind Atari player exploits constraints in its problem domain:
\begin{quotation}
	in the Atari game system, data is very cheap. You can play the game over and over again...get gigabytes of data very quickly, with no real cost.
\end{quotation}
\begin{quotation}
	in the Atari system, ...you have [only] eighteen choices at any given moment.
\end{quotation}
He contrasts this with learning problems of unconstrained choices and sparse data, e.g., when trying to learn from a human or a real-world scenario. Computer games may never be as noisy and unconstrained as real-world scenarios, modern games with multiplayer interaction are frequently very complex. Although \cite{Mnih2015} used games for humans, they were nevertheless from an era of much greater technological constraint, such that their dimensional reduction to tractability was straightforward - modern games will not allow that.

Thus the field of computational intelligence in games faces a research question which I state as \textsf{RQ1}: \textit{how can general game playing AI cope with human-level games and human players?} 

Both \cite{Silver2016,Mnih2015} imposed well-chosen constraints on the problem domain to enable their solutions. Constraints can be dimensionality reduction, and can also be \textit{simulation} of the original system, according to some simplifying theory. 

As stated above, I propose that a generalised player model gives a partial solution. There are two parts to the solution: \textsf{a}) to capture information about player psychology (cognition, emotion and personality) and activity; \textsf{b}) to represent that information in the context of the game. Part \textsf{a}) \textit{constrains} the model of player behaviour to well-understood theoretical constructs; part \textsf{b}) presents the model as input to a learning algorithm.

\subsection{Behavlets}
\label{behav}
A general model should provide insight into different facets of player behaviour, for example the cognitive information processing 'style' of a player. It thus requires a foundation of parameters that describe the subjective experience of play. The foundation will draw on established modelling tools, including at least: i) psychology of behaviour; ii) general game design; and iii) actions in the context of a given game.

I previously proposed the \textit{Behavlets} method \cite{Cowley2016behavlet} to build facets i) to iii) above into composite features of game-play defined over entire action sequences. The aim is to create player-modelling features linked to valid psychological theory. The Behavlet process integrates descriptive models for temperament theory, game design patterns, and patterns of player actions. The core concept is to capture behaviours with certain known bias of personality; e.g. aggression, caution; and thus observe the players' self-expression. Behavlets have been used to model players for, e.g. personality type classification \cite{Cowley2013} and move prediction \cite{Cowley2016:pre2}. Thus I use the Behavlets method to fulfil part \textsf{a}) above.

\subsection{Formalism}
\label{formalism}
How best to represent Behavlets (or any other psychological model) 'in the game', i.e. in a manner both machine- and human-comprehensible? 
In principle, this should be done by \textit{simulation}. As stated in the inspirational work of \cite{Grunvogel2005}, simulation "models' main purposes are to leave out certain aspect of complex systems to facilitate study of those systems."

Note that games can be neatly modelled as a mathematical system because they rely on rule-based interactions defined on a possibility space, and the mechanics of play are essentially functions over that space. 

Restricting the games under consideration to those with strictly bounded rules, observe that a state at time $t$ is determined by the game state at time $t-1$. Thus the game can usually be represented by a finite-state Markov process \footnote{With a non-rational learning human player at the core of gameplay (who may display high \textit{choice variance}, i.e. infer different predicates based on the same observations), game processes are usually strictly non-Markovian; however they can still be given Markovian representations as a simplifying assumption.}. A state-based model is often used for game representation, and Markov methods are often used for computational intelligence in games. 

However, observe that play involves spaces and control systems; these can be either discrete, or approximately continuous with minimum lower bound, sometimes defined by the frame rate of e.g. 60fps or 16.67ms per frame.

For purpose of player modelling, the difference between approximate and truly continuous is not as important as the player's \textit{understanding} of the nature of the play space. To create a general player model we must capture the player's understanding, and deal with 'approximately continuous' data.

If the model must capture every frame of the game, it is hardly an efficient simulation. Far more parsimonious to use a modelling framework that can handle continuous entities. 

For example, consider Go played with clocks. Players make a single discrete move while their clock elapses continuous time. The elapsed time value can be captured with a simple integer, but the elapsed psychological experience cannot.

Fortunately, the required tools are already in \cite{Tabuada2004}'s category theory framework to model interactive control systems. The framework in \cite{Tabuada2004} models both discrete and continuous control systems, in hybrid form and as abstraction simulations. I will draw on the definition of hybrid control systems (HCS), following \cite{Grunvogel2005} and building on \cite{Tabuada2004}. 

\cite{Grunvogel2005} is an excellent complement for the reader; it works lucidly through the foundational technical aspects of applying this formalism to games. It also concludes at about the point where I aim to depart: the composition of micro-games (e.g. Behavlets) to form complete games (e.g. player models). 
The approach is more applied than \cite{Grunvogel2005}, but as in that paper I still aim to produce a simulation model with reduced complexity compared to the original game.

\cite{Grunvogel2005} described the \textit{how} of game specification using HCS methods, but he himself questioned \textit{why} one would wish to do it. I am interested in providing this motivating vision.

\subsection{Summary}
In this paper I aim to provide a notation to represent Behavlets as action sequences in a formally defined simulation of a game system, by extending \cite{Tabuada2004}. The motivation is to generate a representation of possible player actions, and the archetypal behaviour traits that can shape those actions, such that the representation can be used as input for a learning system. Ultimately, the goal is to learn from real human behaviour.

In the rest of the paper, I first give a brief literature review in the next section. In section \ref{formal} I describe a formal model of a game system, before showing briefly in section \ref{disc} how it can be used to represent some Behavlets taken from \cite{Cowley2016behavlet}. Finally, in section \ref{conc}, I suggest some future directions of work.


\section{Background}
\label{bkgd}
A general player model has the difficult task to account for the variation between players, variability in their behaviour over time, and the reciprocal relationship of players to the game. For example, such a model should account not only for player learning, but also player emotions' impact on play. There are many relevant fields of study in that problem, and I have previously reviewed literature contributing to generalised player modelling \cite{Cowley2016:pre1}. Here I briefly review literature on formal models. 

Various \emph{descriptive} models of game play have tried to include aspects of player psychology, such as emotions.
For example, I proposed the User-System-Experience (USE) model \cite{Cowley2006,Cowley2008}, to describe the intrinsic motivation of games in terms of the cognitive neuroscience of information processing and learning. However the specification of games themselves was lacking in detail.
J\"{a}rvinen \cite[pp.99-247]{Jarvinen2009} built a player experience model on top of a game decomposition theory. The model has two concepts: game experiences are composed of sequences of emotions; and game elements embody conditions that elicit emotions.
\cite{Gmytrasiewicz2000} define a formalisation of ‘synthetic' emotions using Decision Theory, to be used for player modelling or for communication of AI agent states to the player. 
Methods which codify game mechanics allow a model to capture player-game interactions. \cite{Sicart2008} attempts this, using the object oriented programming paradigm to define game mechanics as "methods invoked by agents".
\cite{Breining2011} developed a formal modelling toolset to analyse player behaviour by action sequence mining. The method finds all action sequences and their frequency in a game log, representing common sequences as features, which are selected by ranking according to their mutual information with the class variable.

Formal specification of the play space can support the integration of game and psychological models. \cite{VonNeuman1944} defined game theory, which gives useful tools to analyse player behaviour: assuming that players are rational agents with definable utilities for action. Such assumptions do not serve our purpose to learn from real human behaviour. More generally, formal methods such as category theory \cite{Walters1991}, enable specification and verification of the objects and actions of the play space, and thus support rigorous testing of system coherence.
Category theory was applied to game specification in \cite{Grunvogel2005}, which leveraged \cite{Tabuada2004}'s system of notation for abstractions. In \cite{Grunvogel2005}'s abstract specification, a game "consists of objects which change their state during the play, where the evolution of their state is governed by rules and influenced by the players or other objects". \cite{Grunvogel2005} defined a game as a $triple (S, \M, F)$, where $S$ is a set of game states; $\M$ is a monoid describing the inputs to the system; and $F$ is an action of the monoid on the set, i.e. the rules. \cite{Grunvogel2005} also showed how the operation of composition defined in \cite{Tabuada2004} could be used to create novel games; this was a useful abstract discussion. 

This approach is flexible, but the complexity of the domain poses a large problem for this method. \cite{Grunvogel2005} agrees: "describing a game with this formalism seems to be a cumbersome task".
The task is cumbersome because the approach relies too much on one system; any such system will be either unwieldy or insufficiently descriptive. In a multi-step modelling approach, methods for action-tracking \cite{Breining2011}, design pattern analysis \cite{Bjork2005}, and player psychology profiling \cite{Cowley2016behavlet} can first describe the game; i.e. part \textsf{a}) above. These descriptions can then be associated with a coding formalism for rigour, i.e. part \textsf{b}) above.


\section{Formal Model}
\label{formal}
Here, I extend the full HCS defined by \cite{Tabuada2004}.
A game is modelled as a HCS; some rectifying operations are also defined, to force the HCS to behave as games do.

\subsection{Model Foundation}
\label{def1}

\paragraph*{Definition} A game $G=(\X, \M, \Phi)$, consisting of:
\begin{itemize}
	\item the state space $\X=\{\X_q\}_{q \in Q}$ \vspace{0.1cm}

	\item a monoid $\M=\coprod_{n \in \N}(U^*\cup\Sigma^*)^n$ \vspace{0.1cm}

	\item a partial action $\Phi$ of $\M$ on $\X$, such that there exist invariants $Inv(q) \subseteq \X_q$ \vspace{0.1cm}
\end{itemize}

\paragraph*{Note} Here, $\X$ is a set of smooth manifolds parametrised by discrete states $q\in Q$; this allows modelling of any simulated spaces with entities, such as a game's 3D environment with typical player-controlled unit(s) and opponent(s).

\paragraph*{Note} Monoid $\M$ is defined as the product union of the sets $U^*$, the set of smooth manifold inputs, and $\Sigma^*$, the set of discrete inputs \footnote{Although continuous systems are constrained to have finite duration of input times, they may have infinite number of inputs defined as vector field maps from an input manifold. This permits a model consistent with the player's point of view, which is an important part of creating psychologically relevant models.}; which allows modelling of combined analogue and digital inputs, such as a joystick and buttons. Individual inputs are denoted by $m$, a map in $\N^+_0$, defined as a composition of finite $u^{t{1 \dots i}}$ with finite $\sigma_{1 \dots j}$. In this system, $u^{t'}$ indexes time, with 'embedded' discrete inputs from $\Sigma^*$, if modelling game time is required.

\paragraph*{Note} The partial action $\Phi$ implies a ruleset that can be defined over a subset of the state space; this allows modelling of rules such as power-ups, which alter some core function in a restricted area of state-space, i.e. after a power-up item has been consumed, and perhaps within limited time/space.

This general-form model may be revised to obtain the core framework for specific games. For example perfect-information purely-discrete games, such as chess and Go, can be obtained when $\X_q$ is a singleton and $U=\emptyset$. 

\paragraph*{Example}
Let us model the game of \textit{Noughts \& Crosses} (TicTacToe in American) as a demonstration.
\begin{itemize}
	\item $X_{\texttt{xo}} = (pos_\psi)_{\{\psi \in 1 \dots 9 \}}$, the set of $3 \times 3$ board positions, uniquely ordered by the magic square $n=3$ \footnote{This is a rare occasion when a magic square becomes a magic circle (in the sense of Huizinga, not Yang Hui)!}.
	
	\item $\M_{\texttt{xo}} = \sigma_{\texttt{x} \in \psi} \cup \sigma_{\texttt{o} \in \psi}$, the act of placing an $\texttt{x}$ or an $\texttt{o}$.
	
	\item $\Phi_{\texttt{xo}} = \phi : \{1, 2, 3\}$, a map to three 'rules',
	\begin{enumerate}
		\item $\sigma_\texttt{x} \times \sigma_\texttt{o} \longrightarrow {_4P_\psi} $, paired player turns involve sampling without replacement from the magic square $n=3$, up to four times,
		
		\item $\texttt{x} \cap \texttt{o} = \emptyset$, choices are disjoint,
		
		\item $win \iff \sum\Sigma^*=15$, winning condition such that player wins if and only if 3 choices sum to 15.
		
	\end{enumerate}
	
\end{itemize}

Although \textit{Noughts \& Crosses} is a trivial child's game, it is a simple matter to adapt this specification to model \textit{Gomoku}, which is also an $m,n,k$-game. From there, it is straightforward to model Go, at least for $(\X, \M)$. To define $\Phi$ for the core Go rules, which we will not state here for the sake of brevity, would require significant effort but tractable complexity because the rules are all simply derived from the board and input definitions $\X, \M$.

In order to more flexibly create games, it helps to exploit modularity; for this we can use the operation \textit{composition of monoids}. Composition implies that, given two monoids $\M_1$ and $\M_2$, we can form the composition $\M' = \M_1 \otimes \M_2$, which is also a monoid. $\M'$ has all possible evolutions of the composed monoids and no interaction between their parts.

For such a model $(\X, \M, \Phi)$as described, a common shorthand notation is $\Phi_\X$, denoting $\Phi_\X: \X\times \M_x \to \X$. With this notation, and composition, we can thus describe a basic game, $\Phi_0$, and compatible game-parts $\Phi_{\X a}$ and $\Phi_{\X b}$, and obtain a complete game by composition, $\Phi_{\X b} \times \Phi_{\X a} \times \Phi_{\X 0} \rightarrow \Phi_{\X ab}$. The goal is that such game-parts are used to represent Behavlets, as described below, section~\ref{behavlets}.

However as \cite{Grunvogel2005} pointed out, with such a framework it is not yet possible to build any reasonably interesting game, in the sense of  a system which produces meaningful decisions and outcomes \cite{Salen2004}. This is because the composition operator does not impose any interaction on the composed parts, leaving the resultant system causally heterogeneous and un-gamelike. Composition should additionally impose constraints on the composed monoids, such that the inputs of each are influenced by the other. Additionally, tracking activity patterns allows us to see more clearly how the defined influences work in practice.
Therefore, two more concepts will complete our core toolset: \textbf{composition with restriction}, and \textbf{orbits}.

\paragraph*{Definition} \textbf{Composition with restriction}, denoted $\otimes$, from \cite{Tabuada2004}, imposes a restriction of $\Phi_\X$ to a subset of $\X \otimes \M'_x$, such that the composed monoids are forced to synchronise by the restriction.

\paragraph*{Definition} An \textbf{orbit} is a set $O_x$ containing all points visited on an evolution starting at $x$ and controlled by some input $m \in \M$. 
Formally, $O_x = \{x' \in \X : x' = \Phi_\X(x,m')$ for some prefix $m'$ of $m\}$. \cite{Walters1991} defines an orbit as the behaviour of an imperative program $f$, i.e. the effect of a series of inputs $a$ on an initial state $x$, such that $f_{ai}(x_{i-1})=x_i$.  

For our purposes, an orbit of monoid $\Phi_\X$ will represent instances whenever the game-play activity pattern defined by $\Phi_\X$ is played. 
For example, in Noughts \& Crosses there are well-known tactics, which when played according to the correct selection criteria will generate a perfect game. These include \textit{Play Center}, \textit{Block}, \textit{Fork} and others defined in \cite{Crowley1993}. They can be modelled with a triplet defining: the player's move, the state of the board, plus a test to determine the type of tactic played.

In the game of Go there are also various well-known patterns of play, such as \textit{atari}, \textit{gote} vs. \textit{sente}, \textit{joseki}, \textit{ko} fighting. These concepts may be captured by orbits, at least where the analysis of the pattern characteristics is algorithmic and tractable (in Go, an element of expert judgement is often involved in assessing such patterns).

To make a well-formed map from games to models, an orbit for modelling behaviour is under two constraints. It cannot be a cycle, as cycles are not game-like: consider the \textit{ko} rule in Go. It cannot consist only of a stable state, as this cannot evolve (by definition \cite{Walters1991}), and is therefore uninteresting from a player modelling point of view.

\subsection{Complete Model}
\label{def2}
Based on this framework I propose a modelling scheme for game behaviour, which is descriptive rather than generative; i.e. the aim is to simulate the game components that relate to player actions, rather than deriving a simulation of game play from a model of the engine mechanics.

\paragraph*{Definition} A game $G'$ is a composition of HCSs $\Phi_{\X i}$, ${1 < i \leq k}$, where each $\Phi_{\X i}$ is used to model a distinct game play pattern, and the composition (by the properties of composition of monoids) is also a HCS.

The 'base' monoid $\Phi_{\X0}$ represents the core game framework, with no orbit restrictions. A single game play pattern is represented by a monoid, $\Phi_{\X i} : (\X_i, \M_i, \Phi_i)$, instantiated by an orbit $O_{\X i}$, with a starting condition $(q,x_0)_i$, an initial condition from $\X_{qi}$ which corresponds to the opening state of the game pattern. Such monoids are constrained from having initial or terminal objects (as defined by \cite{Walters1991}), because they would then be allowed to define only a single function, violating the principle that games should be uncertain.

Modelling of the complete game is achieved by composition with restriction, where three restriction operators are defined.
\begin{enumerate}
	\item $\X_{qi} \subseteq \X_{qi-1}$, i.e. state space is reduced every time by composition. This models the progression of games, i.e. the fact that a game can generally be modelled as a tree traversal, such that every move will reduce the remaining possible moves.
	
	\item $m_i \otimes m_{i-1}$ iff $m_{i-1}$ is a prefix of $m_i$, such that e.g. in a time-indexed system $m_{t^1} \geq {m-1}_{t^1}$, i.e. the start time of the orbit for the next monoid to be composed must be greater than or equal to the prior monoid, such that game progression is modelled. 
	
	\item $O_{xi} \neq O_{xi-1}$ where $(q,x_0)_i \cap (q,x_0)_{i-1} \neq \emptyset$ without any other restriction on $x \in O_{\X i}$. I.e. monoids composed such that their orbits have overlapping time sequences, shall not be isomorphic.
	
\end{enumerate}

Thus, based on this approach, a game $G'$ is a basis monoid $\Phi_{\X0} : (\X_0, \M_0, \Phi_0)$, which provides the complete state space and time registration, without inputs. The basis monoid is composed with $1..k-1$ additional game pattern monoids, to describe those activities in the game that reduce the possibility space until game end. Each pattern monoid $\Phi_i$ is restricted to join the base monoid in a time-ordered fashion, without overlap of isomorphs, and without expanding the game tree with nodes excluded by previously composed monoids $\Phi_{1..i-1}$.
 


\section{Behavlets-based formal model}
\label{behavlets}

As mentioned, the intention is to formalise the Behavlets' method of player modelling. 
As indicated in the formal model definition, this can be done by using orbits to represent the play patterns which arise in the game. Here I elaborate this idea. 

A Behavlet is essentially a game play pattern associated with a temperament trait. Thus, a well-chosen monoid representation of a pattern, $\Phi_{\X i}$, can also represent a Behavlet, and thereby be associated with a temperament trait. 
To select the right monoid for the Behavlet is quite straightforward. 
The instantiation orbit $O_{\X i}$ is equivalent to the Behavlet logic, defined in \cite{Cowley2016behavlet}. 
Further, the orbit starting condition $(q,x_0)_i$ is equivalent to the Behavlet concept of a \textit{constraint harness}, defined in \cite{Cowley2016behavlet}.

An example will illustrate the approach, for which I will take already peer-reviewed \cite{Cowley2016behavlet} and empirically tested \cite{Cowley2016:pre2} Behavlets, derived for the game \textit{Pac-Man } (Namco, 1980). The formal framework for the Behavlets model of a given game $\Gamma$ is obtained and used with the following five step process:

\begin{enumerate}
	\item define basis monoid for $\Gamma$
	
	\item define compositional $\Gamma$ play pattern monoids, each with associated temperament trait
	
	\item define model \textit{instance} as label for play personality in $\Gamma$
	
	\item model reduction by the operation of \textit{simulation}, giving representations of behaviour patterns in $\Gamma$-like games which can be compared
	
	\item obtain generalised player model by iterating this process
\end{enumerate}

I will apply this process to the Pac-Man Behavlets using the Pac-Man specification from my previous work (see e.g. appendix D to \cite{Cowley2013}). This specification was totally state-based, abstracting the smooth movement aspect of original Pac-Man. Thus, as with chess or Go, this Pac-Man does not need $U^*$, and $\X$ is singleton. For brevity, I will make a number of further simplifications which do not relate to the example Behavlet. I model only a single level, to avoid extra complications surrounding $tests$ that would be needed to model end-of-level or loss of lives (for a description of test function construction, see e.g. \cite[pp.46]{Walters1991}). I do not provide extra state variables to model the bonus Fruit item, a cherry; or record the points scored, or Pac-Man's lives (these values are referenced but left undefined). I also refrain from modelling any driver of Ghost behaviour, which in the prior specification is simply a probabilistic map to adjacent positions, weighted toward Pac-Man in normal play and away from Pac-Man when a power pill is in effect. All these features can be trivially added to the model.

\vspace*{1em}
First, the basis monoid for a Pac-Man game $\PM$.

\paragraph*{Definition} $\Phi_\PM = (\X_{\PM0}, \M_{\PM0}, \Phi_{\PM0})$,
\begin{itemize}
	\item $\X_{\PM0} = \{mat, xy_p\}$, where $mat = (hpos_x) \cup (vpos_y), \{x,y \in \mathbb{N}, 1 \dots 20\}$ the Pac-Man 'map' matrix with values drawn from $\{\emptyset, wall, pill, powerpill\}$; and $xy_p$ is a set of current position values for Pac-Man and the Ghosts $\{ xy_p ~\vert~ {p \in PM, G1..4}\}$,
	
	\item $\M_{\PM0} = m : \{\leftarrow, \uparrow, \downarrow, \rightarrow\} \times \X_{xy_{PM}} \xrightarrow{d=1} \X$, a map from the four directions of movement to the matrix position adjacent to Pac-Man's current position,
	
	\item $\Phi_{\PM0} = \phi : \{1, 2, 3\}$, a map to three 'rules',
	\begin{enumerate}
		\item $m \times pill \longrightarrow \{+5_{points}, \X_{xy_{PM}} = \emptyset\}$, Pac-Man passes through a matrix position with a pill: increase points by +5, position becomes empty, 
		
		\item $m \times powerpill \longrightarrow \{+10 points, \X_{xy_{PM}} = \emptyset, \phi = \phi' : \{1, 2, 3'\}, t : (1..n) \}$, similar effects as $pill$; also transition to the map $\phi'$, where vulnerability of Pac-Man to the Ghosts is inverted for a limited time $n$, 
		
		\item $m \times xy_{G1..4} \longrightarrow -1_{life}$, Pac-Man and a Ghost enter the same matrix position: Pac-Man loses a life,
		
		\item[$3'$] $m \times xy_{G1..4} \longrightarrow \{+50points \times i, i \in (1..4), xy_{G_i} = xy_{0G_i}\}$, Pac-Man and a Ghost enter the same matrix position: +50 points (multiplied by consecutive Ghost order), Ghost returns to starting position
		
	\end{enumerate}
	
\end{itemize}

\vspace*{1em}
Second, the Behavlets themselves are modelled. To illustrate, I select a Behavlet listed in \cite[pp.293]{Cowley2016behavlet}, \textit{A1\_Hunt Close To Ghost House}. Behavlet \textit{A1} (for short) tracks how often a player follows the Ghosts right up to their house while attacking them in $powerpill$ mode.

\paragraph*{Definition} $\Phi_{A1} = (x', m', O_{\PM A1})$,

\begin{itemize}
	\item $x' = \forall 1..4, \texttt{dist}(xy_{G_i}, xy_{0G_i}) \leq 3$, the manhattan distance of each Ghost to its own starting position is three or less,
	
	\item $m' \subseteq \M_{\PM0} \forall t_{(1..n)}$, the orbit elapses for all inputs until the end of the $powerpill$ timer,
	
	\item $O_{\PM A1} = \{x' \in \X_{\PM0} : x' = \Phi_{\PM0}(x,m')$, an orbit defined on the Pac-Man basis monoid
	
\end{itemize}

\vspace*{1em}
Third, the composited game is produced, $\PM = \Phi_{A1} \otimes \Phi_\PM = \Phi_{\PM A1}$. 
Game instances where this Behavlet monoid appears can be labelled as examples of \textit{cautious play}, with a quantification scheme as described above. 

\vspace*{1em}
Fourth, model reduction by simulation creates a simpler representation without reference to the specifics of the game. Thus we do not need to define, for example, the dimensions of game space states $\X_q$, only to define the \textit{type} of spaces as they appear to the player. In this way, we can make an equivalence between models for Pac-Man, Go, even Noughts \& Crosses if we wish. Given the game produced by composition with restrictions $\Phi_{\PM A1}$, we define two simulations of the composed parts, $\beta_{A1}$ and $\beta_\PM$. The map $\varphi$ which defines each $\beta$ is an abstraction of the part of the model \textit{which is not relevant to comparison with another simulated game}. $\beta$s are composed by restriction to produce a more general version of the model, $\beta_{\PM A1}$.

\begin{equation}
	\begin{tikzcd}[column sep=tiny]
		\beta_{A1} 			                 & \otimes & \beta_\PM \arrow{r} &[4em] \beta_{\PM A1} \\
		\Phi_{A1} \arrow{u}{\varphi} & \otimes & \Phi_\PM \arrow{u}{\varphi} \arrow{r} & \Phi_{\PM A1} \arrow{u}
	\end{tikzcd}
\end{equation}


The complete approach to simulation is detailed in \cite{Tabuada2004}, and is also discussed in \cite{Grunvogel2005}. Here it is enough to note that, given proven models of games such as those described above, the reduction can be pursued and the simulations are then 'safe' to study without reference to the messy details.

\vspace*{1em}
Fifth and finally, obtaining the generalised player model from the given framework is perhaps possible for a class of games between which simulation is well-defined. Proving this is clearly a matter for future work.


\section{Discussion}
\label{disc}
The approach I described for a generalised player model draws on the Behavlet method to create psychologically-based features of game play, and redefines them as parts of a category theoretic formal model. The value of this approach is that, under a formal framework, Behavlet models of particular games can be further generalised by the operation of simulation.

\subsection{Potential applications}
The primary use case for the described method is to capture player variation. Consider that, if we model Behavlets as game parts $\Phi_\X$, then by the Behavlets method \cite{Cowley2016behavlet}, each $\Phi_{\X i}$ will have an associated behaviour trait. Thus, a game instance with a specific $\Phi$ composition will reflect the 'character' of a particular player's play style. Potentially, characteristics of human play can be learned through enough such instances.

Also consider that the method allow abstraction and specificity: we can build a canonical game model and also simulate game instances quite easily from the same definitions. This allows exploration of the space of possible games.

This might be termed personality profiling, but skill and strategy are also a relevant considerations. Based on the work of \cite{Acuna2010} where humans achieved near-optimal solutions on hard problems, we can expect to find many problems where human skills can provide the seed for improved computational solutions. For example, \cite{Sorensen2016} used a gamification to learn from humans solutions, creating heuristic optimisation methods for quantum computing problems which outperform traditional multi-parameter numerical methods. The problem remains to characterise player activity in a manner which is flexible to the level and type of detail required, a problem to which I suggest that simulation is well suited. 
Thus, given a composite model of Behavlet-based game parts, defining play-sequences such as the Noughts \& Crosses or Go tactics described, the actually-played components can denote the level of insight of the player into the game problem.

A further consequence of this flexibility is the capacity to model multiple \textit{flavours} of game rules. As defined by \cite{Salen2004}, a game contains three different types of rule, Constituative (written before play), Operative (emergent during play) and Implicit (unspoken 'house' rules between players). Go is an example where all these rule types have been studied, standardised, and written about in great detail, and thus were available to the developers of alphaGo. For less well-studied games, the method I present can characterise them with some flexibility without loss of rigour.

\subsection{Issues and considerations}
This is a work at the concept stage, and like any concept there are many details lacking. The state of the method presented is probably sub-optimal, and this may frustrate the more engineering-minded reader; but the aim is initiate a conversation. It is to be hoped that the concept will provide fertile soil to grow more detailed methods.

Despite the seeming complexity, what is required to build such models is quite complementary to the development process - defining the entities and operations of game-play.

When building such models, the user must take note of whether the restrictions and constraints ever contradict his game: this can help highlight flaws in either the game or the model, and facilitate the work of quality assurance.

\section{Conclusion}
\label{conc}
I argue that in order to advance general game playing AI, it is necessary to include the player perspective, because games are ultimately human artefacts and therefore contain cases which benefit from a human-style problem solving approach. The argument implies creating a generalised player model. I have set out a method to do so, based on integrating my previously published Behavlets work \cite{Cowley2016behavlet} with a formal model of game play. The result is a \textit{vision} for a general player model, rather than a complete and final work, which I hope will serve as inspiration.


\bibliographystyle{IEEEtran}
\bibliography{Cowley_CIG16_bib}


\end{document}